\documentclass[11pt,a4paper]{article}
\usepackage{epsfig}
\def\nl{\nonumber\\}
\def\beq{\begin{equation}}
\def\eeq{\end{equation}}
\def\beqar{\begin{eqnarray}}
\def\eeqar{\end{eqnarray}}
\def\barr#1{\begin{array}{#1}}
\def\earr{\end{array}}
\def\bfi{\begin{figure}}
\def\efi{\end{figure}}
\def\btab{\begin{table}}
\def\etab{\end{table}}
\def\bce{\begin{center}}
\def\ece{\end{center}}

\def\text{\textstyle}

\def\la{\lambda}
\def\si{\sigma}

\def\refeq#1{\mbox{(\ref{#1})}}

\def\citere#1{\mbox{Ref.~\cite{#1}}}
\def\citeres#1{\mbox{Refs.~\cite{#1}}}

\newcommand{\GeV}{\unskip\,\mathrm{GeV}}

\newcommand{\TeV}{\unskip\,\mathrm{TeV}}
\newcommand{\fba}{\unskip\,\mathrm{fb}}

\def\mathswitchr#1{\relax\ifmmode{\mathrm{#1}}\else$\mathrm{#1}$\fi}
\newcommand{\PW}{\mathswitchr W}

\def\mathswitch#1{\relax\ifmmode#1\else$#1$\fi}

\newcommand{\scrs}{\scriptscriptstyle}
\newcommand{\sw}{\mathswitch {s_{\scrs\PW}}}
\newcommand{\cw}{\mathswitch {c_{\scrs\PW}}}
\newcommand{\rw}{{\mathrm{W}}}

\newcommand{\cew}{C^{\ew}}

\def\ie{i.e.\ }

\newcommand{\ord}{{\cal O}}

\newcommand{\rR}{\mathrm{R}}
\newcommand{\rL}{\mathrm{L}}
\newcommand{\rT}{{\mathrm{T}}}
\newcommand{\rS}{{\mathrm{S}}}
\newcommand{\rd}{{\mathrm{d}}}
\newcommand{\rc}{{\mathrm{c}}}

\newcommand{\pT}{p_{\mathrm{T}}}
\newcommand{\pTcut}{p_{\mathrm{T}}^{\mathrm{cut}}}
\newcommand{\ew}{\mathrm{ew}}
\newcommand{\M}{{\cal {M}}}

\newcommand{\logar}[2]{\mathrm{L}^{#1}_{#2}}

\newcommand{\LL}{\mathrm{LL}}
\newcommand{\shat}{{\hat s}}
\newcommand{\that}{{\hat t}}
\newcommand{\uhat}{{\hat u}}
\newcommand{\rhat}{{\hat r}}

\newcommand{\pdf}[4]{f_{#1,#2}(#3)}

\newcommand{\A}[2]{A^{(#1)}}

\makeatletter

\def\eqnarray{\stepcounter{equation}\let\@currentlabel=\theequation
\global\@eqnswtrue
\global\@eqcnt\z@\tabskip\@centering\let\\=\@eqncr
$$\halign to \displaywidth\bgroup\hskip\@centering
  $\displaystyle\tabskip\z@{##}$\@eqnsel&\global\@eqcnt\@ne
  \hskip 2\arraycolsep \hfil${##}$\hfil
  &\global\@eqcnt\tw@ \hskip 2\arraycolsep $\displaystyle\tabskip\z@{##}$\hfil
   \tabskip\@centering&\llap{##}\tabskip\z@\cr}

\makeatother

\begin{document}

\thispagestyle{empty} 

\thispagestyle{empty}
\def\thefootnote{\fnsymbol{footnote}}
\setcounter{footnote}{1}
\null
\hfill   TTP04-18\\
\strut\hfill  SFB/CPP-04-32\\
\strut\hfill hep-ph/0408308
\vskip 0cm
\vfill
\begin{center}
{\Large \bf
Logarithmic electroweak corrections to hadronic 
$Z+1$ jet production at large transverse momentum
\par}
\vskip 1em
{\large
{\sc Johann~H.~K\"uhn\footnote{Johann.Kuehn@physik.uni-karlsruhe.de}, 
A.~Kulesza\footnote{ania@particle.uni-karlsruhe.de},
S.~Pozzorini\footnote{pozzorin@particle.uni-karlsruhe.de},
M.~Schulze\footnote{schulze@particle.uni-karlsruhe.de} }}
\\[.5cm]
{\it Institut f\"ur Theoretische Teilchenphysik, 
Universit\"at Karlsruhe \\
D-76128 Karlsruhe, Germany}
\par
\end{center}\par
\vfill 
{\bf Abstract:} \par 

We consider hadronic production of a $Z$ boson in association with a jet  
and study one- and two-loop electroweak logarithmic corrections 
in the region of high $Z$-boson transverse momentum, $\pT\gg  M_Z$,
including leading and next-to-leading logarithms. Numerical results for the LHC and Tevatron colliders are presented.
At the LHC these corrections amount to tens of per cent and will be important for interpretation of the measurements.


\par
\vskip 1cm
\noindent
August 2004 
\par
\null
\setcounter{page}{0}
\clearpage
\def\thefootnote{\arabic{footnote}}
\setcounter{footnote}{0}

\newpage

\section{Introduction}

Hadronic weak-boson production in association with jets forms one of the 
most important classes of Standard Model processes. 
Due to its relatively high cross section and clean experimental signature 
the $Z+1$ jet production process 
can be used for a precise determination of the parton
distribution functions \cite{Catani:2000jh}.
In particular, the precise measurement of the $Z$-boson transverse momentum
($\pT$) distribution combined with theoretical predictions of correspondingly 
high accuracy will permit to extract the gluon distribution function
with unprecendented accuracy at the Large Hadron Collider (LHC).
In order to achieve desired precision (at the per cent level)
radiative corrections should be included in the theoretical predictions.

The next-to-leading order QCD corrections to the total cross section and  
$\pT$ distribution for hadronic $Z+1$ jet  production have been known for a relatively long time by now~\cite{Ellis:1981hk,Giele:dj}. Recent
predictions for the Tevatron and the LHC can be found in \citeres{Campbell:2002tg} and~\cite{Campbell:2003hd}, respectively.
The size of these corrections can amount to several tens of per cent
depending on the observable under consideration, including jet definition, as well as the renormalisation and factorisation scales.

On the basis of simple arguments regarding the strength of 
coupling constants one might expect electroweak (EW) corrections to be in
general much smaller than QCD corrections.
However, in the region  $\sqrt{\shat}\gg M_W\simeq M_Z$,
where the center-of-mass energy $\sqrt{\shat}$ of partonic scattering becomes much larger than the weak-boson mass scale,
the EW corrections are strongly enhanced by 
logarithmic mass singularities.
At $\ord(\alpha^L)$ the leading logarithms (LLs) are
of the form $\alpha^L\log^{2L}(\shat/M_W^2)$,
the next-to-leading logarithms (NLLs) are of the form 
$\alpha^L \log^{2L-1}(\shat/M_W^2)$, etc.
These EW logarithms originate from soft/collinear emission
of virtual EW gauge bosons ($\gamma,Z,W$) off initial- or final-state 
particles.
Owing to the finite weak-boson masses,
the real emission of a soft/collinear $Z$ or $W$ boson
can be observed as a separate process and hence does not need to be included 
in the definition of  physical observables.
Thus, in contrast to mass singularities in massless gauge theories such as  QED or QCD, the EW mass singularities of virtual origin are not necessarily compensated by corresponding mass singularities from real weak-boson radiation.

Typically, at $\sqrt{\shat}\simeq 1 \TeV$ the EW logarithms
are estimated to yield one-loop corrections of tens of per cent 
and two-loop corrections of a few per cent.
At one loop it has been proven that 
the EW LLs and NLLs are universal and results 
applicable to arbitrary processes are available 
\cite{Denner:2001jv,Pozzorini:rs}. 
For  $\gamma W$ and $Z W$  production at the LHC
it was shown that these EW logarithmic corrections
 can become of the order of the QCD corrections
 \cite{Accomando:2001fn}. Furthermore, in \citere{Hollik:2004tm}
the EW one-loop corrections to the $\gamma Z$ production process 
at the LHC were found numerically sizeable, with the logarithmic part providing
 a major contribution to the corrections. 
The resummation of the EW logarithms was 
discussed first in \citere{Kuhn:1999de} with emphasis on four-fermion processes
(for an early discussion see also \citere{Ciafaloni:1999ub}). 
General resummation prescriptions for the EW LLs and NLLs were derived 
in \citeres{Fadin:2000bq,Kuhn:2000nn} and 
\citeres{Kuhn:2000nn,Melles:2001gw}, respectively.
For massless four-fermion processes the EW logarithms 
were  resummed up to the NNLLs \cite{Kuhn:2001hz},
and it was found that, at the TeV scale,
the leading and subleading logarithms have similar size 
and alternating signs, which gives rise  
to large cancellations.
These resummation prescriptions were  confirmed by 
various two-loop  calculations at the 
LL  level \cite{Melles:2000ed,Hori:2000tm,Beenakker:2000kb}
and NLL level \cite{Denner:2003wi,Pozzorini:2004rm}.
Only few two-loop results based on a
diagrammatic calculation exist beyond the NLL level. 
In particular,
the corrections to a massless fermionic form factor
within an  abelian massive gauge theory
\cite{Feucht:2003yx,Feucht:2004rp} were  evaluated up to
$\mathrm{N}^4\LL$ approximation and agreement with the earlier NNLL results
was  observed.

In this letter we investigate the $Z+1$ jet production at the
LHC and the Tevatron. In particular, we focus on 
the high-$\pT$ region, $\pT\gg M_Z$, 
and discuss the impact of the one- and two-loop  EW logarithms on the total cross section and the $\pT$ distribution 
for this process. 
We also compare the one-loop logarithmic corrections 
with the results presented in \citere{Maina:2004rb}, where
the one-loop weak corrections were evaluated numerically.

\section{Analytic results}
The hadronic production of a $Z$ boson with finite transverse momentum $\pT$ in association 
with a jet receives contributions from the partonic subprocesses
\beq
q\bar q\to Z g,\qquad
q g \to Z  q,\qquad
\bar q g  \to Z  \bar q.
\eeq
The  $\pT$ distribution for the hadronic process $h_1 h_2 \to Z + j$
reads
\beqar\label{hadroniccs}
\frac{\rd \si^{h_1 h_2}}{\rd \pT}
&=&
\sum_{q=u,d,c,s,b}\int_0^1\rd x_1 \int_0^1\rd x_2
\;\theta(x_1 x_2-\hat\tau_{\rm min})
\Biggl\{
\biggl[
\pdf{h_1}{q}{x_1}{\mu^2}\pdf{h_2}{\bar q}{x_2}{\mu^2}
+(1\leftrightarrow 2) \biggr]
\frac{\rd \hat{\si}^{q\bar q}}{\rd \pT}
\nl&&{}+
\biggl[
\biggl(
\pdf{h_1}{q}{x_1}{\mu^2}\pdf{h_2}{g}{x_2}{\mu^2}
+
\pdf{h_1}{\bar q}{x_1}{\mu^2}\pdf{h_2}{g}{x_2}{\mu^2}
\biggr)
+(1\leftrightarrow 2) \biggr]
\frac{\rd \hat{\si}^{q g}}{\rd \pT}
\Biggr\}
,
\eeqar
where 
 $\pdf{h}{i}{x}{\mu^2}$ are the parton distribution functions, 
$\hat \tau_{\rm min} =(\pT+m_\rT)^2/s$ 
with  $\sqrt{s}$ denoting the collider energy and
$m_{\rT}=\sqrt{\pT^2+ M_Z^2}$.
The $\pT$ distribution 
for the unpolarized partonic subprocess $ij\to Z k$
reads
\beqar\label{partoniccs}
\frac{\rd \hat{\si}^{i j}}{\rd \pT}
&=&
\frac{\pT}{8\pi N_{i j}\shat|\that-\uhat|}
\left[
\overline{\sum_{\mathrm{pol}}}|\M^{i j}|^2+(\that\leftrightarrow \uhat)
\right]
,
\eeqar
where $\shat=(p_i+ p_j)^2$, $\that=(p_i- p_Z)^2$, and $\uhat=(p_j-p_Z)^2$ are the
Mandelstam invariants.
The momenta $p_{i,j,k}$ of the partons are assumed to be massless, whereas 
$p_Z^2= M_Z^2$ for the $Z$-boson momentum. In terms of $x_1,x_2,\pT$, and the collider energy $\sqrt{s}$,
we have
\beq
\shat=x_1 x_2 s,\qquad
\that=\frac{ M_Z^2-\shat}{2}(1-\cos\theta),\qquad
\uhat=\frac{ M_Z^2-\shat}{2}(1+\cos\theta),
\eeq
where $\cos\theta=\sqrt{1- 4\pT^2 \shat/(\shat- M_Z^2)^2}$
corresponds to the cosine of the angle between the momenta $p_i$ and $p_Z$ in the partonic center-of-mass frame.
The factor $1/N_{i j}$ in \refeq{partoniccs}, with $N_{q\bar q}=N_\rc^2$, $N_{q g}=N_\rc (N_\rc^2-1)$, and $N_\rc=3$, accounts for  the initial-state colour average.

The unpolarised squared amplitude for the $q\bar q\to Z g$ process reads 
\beq\label{generalamplitude}
\overline{\sum_{\mathrm{pol}}}|\M^{q\bar q}|^2=
8 \pi^2
{\alpha\alpha_\rS}
(N_\rc^2-1)
\frac{\that^2+\uhat^2+2 M_Z^2 \shat}{\that\uhat} 
\left[
\A{0}{\la}+\left(\frac{\alpha}{2\pi}\right)\A{1}{\la}
+\left(\frac{\alpha}{2\pi}\right)^2 \A{2}{\la}
\right]
,
\eeq
where $\alpha$ and $\alpha_\rS$ are the electromagnetic and 
strong coupling constants, respectively.
The amplitude for the $q g$-initiated  process 
is easily obtained from \refeq{generalamplitude}
using crossing symmetry, 
\beq\label{crossing}
\overline{\sum_{\mathrm{pol}}}|\M^{q g}|^2= -
\left.\overline{\sum_{\mathrm{pol}}}|\M^{q\bar q}|^2 
\right|_{\shat\leftrightarrow \uhat}
.
\eeq
The tree-level contribution $\A{0}{\la}$ to \refeq{generalamplitude} reads
\beq
\A{0}{\la}=\sum_{\la=\rL,\rR}
\left(I^Z_{q_\la}\right)^2
\qquad\mbox{with}\qquad
I^Z_{q_\la}=\frac{\cw}{\sw} T^3_{q_\la}-\frac{\sw}{\cw}\frac{ Y_{q_\la}}{2},
\eeq
where $T^3_{q_\la}$ and $Y_{q_\la}$
are the weak isospin and hypercharge for left- ($\la=\rL$)
and right-handed ($\la=\rR$) quarks, and we use 
the shorthands $\cw=\cos{\theta_\rw}$ and  $\sw=\sin{\theta_\rw}$
for the  weak mixing angle $\theta_\rw$. 

The virtual electroweak corrections were evaluated in the 
high energy region where all invariants are much larger than the weak-boson mass scale,
\beq
|\rhat|\gg  M_W^2\sim  M_Z^2\qquad\mbox{for}\qquad \rhat=\shat,\that,\uhat.
\eeq
In this region, using the results of \citeres{Denner:2001jv,Denner:2003wi,Melles:2001gw}, 
we have computed the one- and two-loop electroweak logarithms
of the type $\logar{k}{\rhat}=\log^k(|\rhat|/ M_W^2)$
to next-to-leading logarithmic accuracy.
At $L$-loop level this approximation includes leading logarithms (LLs) and next-to-leading logarithms (NLLs),
\ie contributions of order 
$\logar{2L}{\shat}$ and $\logar{2L-1}{\shat}$, respectively. 
Contributions of order $\logar{2L-2}{\shat}$ are systematically omitted. 
In particular, for the angular-dependent logarithms, \ie  
logarithms of the type $\logar{2L}{\rhat}$ and  $\logar{2L-1}{\rhat}$ involving invariants $\rhat= \that,\uhat$, we use the approximation
\beq
\logar{2L}{\rhat}=\logar{2L}{\shat}+2L \log\left(\frac{|\rhat|}{\shat}\right) \logar{2L-1}{\shat}
+\ord(\logar{2L-2}{\shat}),\qquad
\logar{2L-1}{\rhat}=\logar{2L-1}{\shat}+\ord(\logar{2L-2}{\shat}).
\eeq

In order to separate the logarithmic electroweak corrections into 
finite and infrared-divergent parts in an elegant way%
\footnote{
Using the infrared-evolution-equation (IREE) \cite{Fadin:2000bq}, 
the electroweak logarithmic corrections can be factorized in 
a SU(2)$\times$U(1)-symmetric and finite part times 
a $\mathrm{U}(1)_{\mathrm{em}}$-symmetric and infrared-divergent part.
The SU(2)$\times$U(1)-symmetric part corresponds to the 
complete electroweak corrections with infrared divergences 
regularized by a fictitious photon mass $\lambda = M_W$.
The $\mathrm{U}(1)_{\mathrm{em}}$-symmetric part contains the infrared divergences 
resulting from the gap
between the (vanishing) photon mass and the weak-boson mass scale. 
Within the framework of the IREE these two parts factorize and 
exponentiate separately.
At the LL level, the correctness of this IREE-approach has been 
completely confirmed by process-independent two-loop electroweak 
calculations based on different gauge-fixings 
\cite{Beenakker:2000kb,Denner:2003wi}.
Recent two-loop calculations \cite{Denner:2003wi,Pozzorini:2004rm,Feucht:2004rp} have shown that the IREE-approach 
is applicable also beyond LL level.
}%
,
it is useful to introduce a fictitious photon mass $\lambda = M_W$.
In the present paper we restrict ourselves to the 
finite part of the corrections which is defined as the 
complete electroweak correction with  $\lambda = M_W$.
The remaining part, which is not considered here,
depends on the gap between the (vanishing) photon mass and the 
weak-boson mass scale and is infrared divergent.
This infrared-divergent part can be factorized and resummed separately
\cite{Fadin:2000bq} and needs to be combined with real photon radiation. 
Its size depends on the choice of cuts on the photon energies and angles.

The small contributions resulting from logarithms of the $Z$-$W$ mass ratio are neglected.
Mass-suppressed corrections of order $ M_W^2/|\rhat|$ are not considered
and, since the contributions from 
longitudinally polarised $Z$ bosons are mass-suppressed in the high energy limit, the logarithmic corrections have been computed assuming that the 
external $Z$ boson is transversely polarised.
At one-loop level we obtain 
\beq\label{oneloopresult}
\A{1}{\la}= - \sum_{\la=\rL,\rR}
I^Z_{q_\la}\left[
I^Z_{q_\la}\cew_{q_\la}\left(\logar{2}{\shat}-3\logar{}{\shat}\right)
+
\frac{\cw}{\sw^3}T^3_{q_\la}
\left(\logar{2}{\that}+\logar{2}{\uhat}-\logar{2}{\shat}\right)
\right]
,
\eeq
where $\cew_{q_\la}=Y_{q_\la}^2/(4\cw^2)+C_{q_\la}/\sw^2$
are the eigenvalues of the electroweak Casimir operator 
for quarks, with $C_{q_\rL}=3/4$  and $C_{q_\rR}=0$. 
The above result takes into account LLs and NLLs of soft/collinear origin
as well as NLLs resulting from the running of the coupling constants.%
\footnote{
We note that in addition to the logarithmic corrections in \refeq{oneloopresult} the partonic subprocesses involving $b$ quarks receive also NLL corrections proportional to $M_t^2/M_W^2$,  which originate from the 
Yukawa interaction in the heavy-quark doublet \cite{Denner:2001jv}.
At partonic level these corrections can be relatively large.
However, their contribution to the hadronic cross section
is almost negligible owing to the smallness of the $b$-quark density of protons and anti-protons.
We have checked that the correction resulting from these Yukawa terms 
does not exceed one per mille at the LHC.} 
The latter NLLs are, however, cancelled by the collinear logarithms 
that are associated with the final-state $Z$ boson \cite{Denner:2001jv}
and thus do not appear in the result \refeq{oneloopresult}.
In the one-loop approximation and for the
process under consideration
the electroweak corrections can be split into electromagnetic and
weak parts in a gauge-invariant way.
The weak corrections (which result from 
virtual $Z$- and $W$-boson exchange) are easily obtained from 
\refeq{oneloopresult} through the  substitution  $\cew_{q_\la}\to
\cew_{q_\la}-Q^2_{q}$, where $Q_{q}$ denotes the electromagnetic
charge of the quark $q$. 

At two-loop level we obtain
\beqar\label{twolooplogs}
\A{2}{\la}&=&
\sum_{\la=\rL,\rR}\Biggl\{
\frac{1}{2}
\left(
I^Z_{q_\la}\cew_{q_\la}
+
\frac{\cw}{\sw^3}T^3_{q_\la}
\right)
\Biggl[
I^Z_{q_\la}\cew_{q_\la}\left(\logar{4}{\shat}-6\logar{3}{\shat}\right)
\nl&&{}
+
\frac{\cw}{\sw^3}T^3_{q_\la}
\left(\logar{4}{\that}+\logar{4}{\uhat}-\logar{4}{\shat}\right) 
\Biggr]
-\frac{T^3_{q_\la}Y_{q_\la}}{8\sw^4}
\left(\logar{4}{\that}+\logar{4}{\uhat}-\logar{4}{\shat}\right) 
\nl&&{}
+\frac{1}{6}I^Z_{q_\la}
\Biggl[I^Z_{q_\la}\left(
\frac{b_1}{\cw^2}\left(\frac{Y_{q_\la}}{2}\right)^2
+\frac{b_2}{\sw^2} C_{q_\la}
\right)
+\frac{\cw}{\sw^3}T^3_{q_\la}b_2
\Biggr]\logar{3}{\shat}
\Biggr\},
\eeqar
where $b_1=-41/(6\cw^2)$ and $b_2=19/(6\sw^2)$ are the one-loop $\beta$-function coefficients associated with the U(1) and SU(2) couplings, respectively.
We note that at the two-loop level the purely weak corrections cannot be isolated from the complete electroweak corrections in a gauge-invariant way.
The LLs as well as the angular-dependent subset of the NLLs in \refeq{twolooplogs}, \ie all contributions of the form $\logar{4}{\rhat}$
with $\rhat=\shat,\that,\uhat$, have been derived from \citere{Denner:2003wi}.
There, by means of a diagrammatic two-loop calculation
in the spontaneously broken electroweak theory, it was shown that such two-loop terms result from the exponentiation of the corresponding one-loop corrections. 
The additional NLLs of the form  $\logar{3}{\shat}$ in \refeq{twolooplogs}
have been obtained via a fixed order expansion
of the general (process independent) resummed expression proposed  in \citere{Melles:2001gw}.
This NLL resummation \cite{Melles:2001gw} relies on the 
assumption that effects from spontaneous breaking of the SU(2)$\times$U(1) symmetry  can be neglected in the high energy limit.

\section{Numerical results and discussion}

In this section we discuss numerical effects of the one-
and two-loop logarithmic EW corrections on the $Z+1$ jet cross sections
at the LHC and the Tevatron. Results presented here are obtained for the 
following values of the electroweak parameters: 
$\alpha=\alpha( M_Z^2)=1/127.9$, 
$\sw^2=0.231$,
$ M_Z=91.19 \GeV$ and $ M_W= M_Z \sqrt{1-\sw^2}$.
When discussing the (relative) statistical error, defined as 
$\Delta \si_{\mathrm{stat}}/\si=1/\sqrt{N}$ with $N={\cal L}\times$BR$(Z\rightarrow l, \nu_l)\times \si_{\mathrm LO} $, 
we include all leptonic decays
of $Z$ with the branching ratio \mbox{BR($Z\rightarrow l, \nu_l) =30.6 \%$} and assume
total integrated luminosity of ${\cal L}=300  \fba^{-1}$
 and  ${\cal L}= 11 \fba^{-1}$ for
  the LHC~\cite{LHClum} and the Tevatron~\cite{TEVlum}, respectively.

To calculate the hadronic cross sections we use the leading order parton
distribution functions MRST2001LO~\cite{Martin:2002dr} at the factorisation
scale $\mu^2_{\rm F}=\pT^2$. Similarly, the strong coupling constant $\alpha_\rS$ is evaluated
(for five active flavours) at the scale $\mu^2 = \pT^2$  and,
consistently with the parton distribution analysis, we use the one-loop
expression with $\alpha_\rS( M_Z^2)=0.13$~\cite{Martin:2002dr}.
 
\begin{figure}[h!t]
 \begin{center}
\epsfig{file=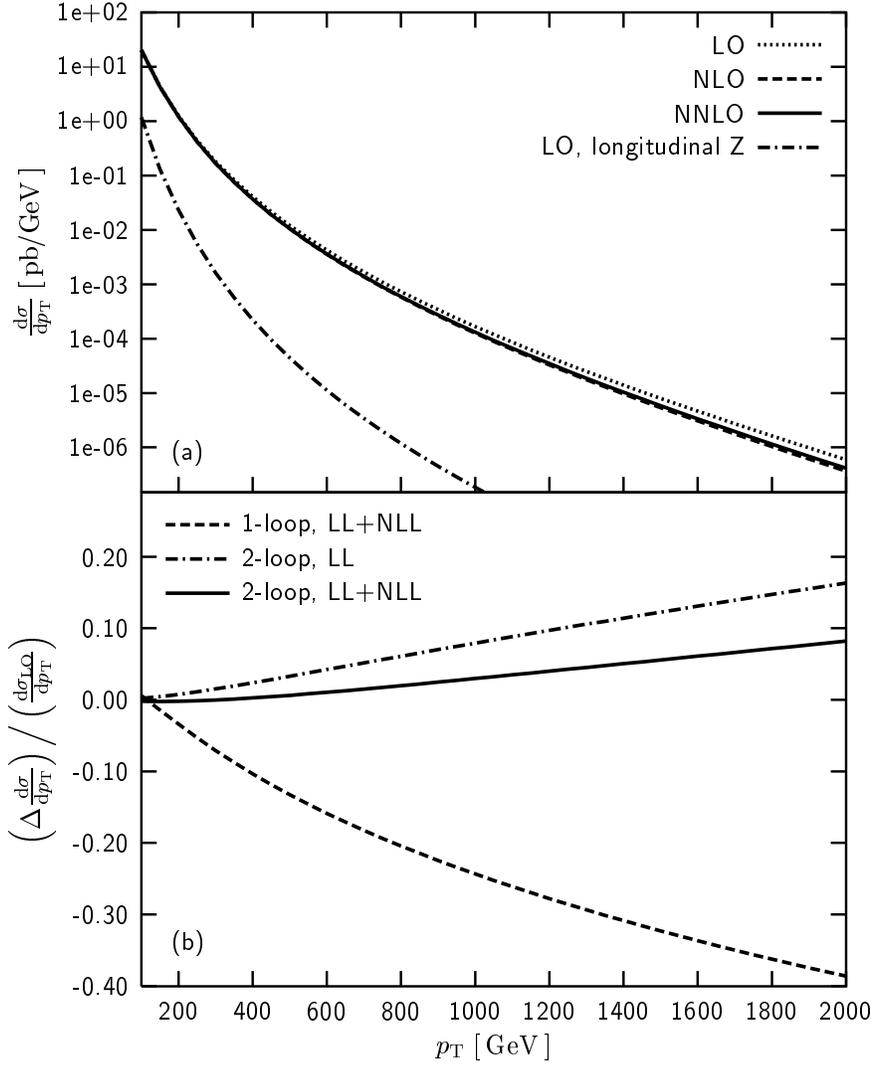, angle=0, width=11.5cm}
\end{center}
\caption{
(a)  Transverse momentum  distribution for $ p p \to Z j$ at $\sqrt{s}=14\TeV$: 
LO (dotted), NLO (dashed) and NNLO (solid) result for unpolarised $Z$ and 
LO contribution from  longitudinally polarised $Z$ (dash-dotted).
(b) 
Relative electroweak correction to the lowest order unpolarised $\pT$ distribution for $ p p \to Z j$ at $\sqrt{s}=14\TeV$:
1-loop LLs+NLLs (dashed), 
2-loop LLs (dash-dotted) and 2-loop LLs+NLLs (solid). 
}
\label{fig:lhc1}
\end{figure}

No rapidity cuts are imposed for the purpose of this study, as their
effect on the $\pT$ distributions does not appear to be relevant, both at the 
LHC and the Tevatron. We checked that at the LHC a rapidity cut
$|y_Z| < 4$ decreases the $\pT$ distribution at $\pT=100\GeV$
only by less than 0.5\% and for $\pT\gg 100\GeV$ the effect of the rapidity cut
becomes negligible.

The lowest order (LO)  transverse momentum
distribution of the outgoing $Z$ boson at the LHC is shown in
Fig.~\ref{fig:lhc1}a. We also display seperately the lowest order 
 contribution from the longitudinally polarised $Z$ boson. 
Its smallness justifies 
to treat the radiative corrections to the unpolarised process
as if the $Z$ boson would be only transversally polarised.
The next-to-leading order (NLO) and next-to-next-to-leading order (NNLO)
results, \ie one- and two-loop corrected distributions, are also
presented.

\begin{figure}
\vspace*{2mm}
  \begin{center}
\epsfig{file=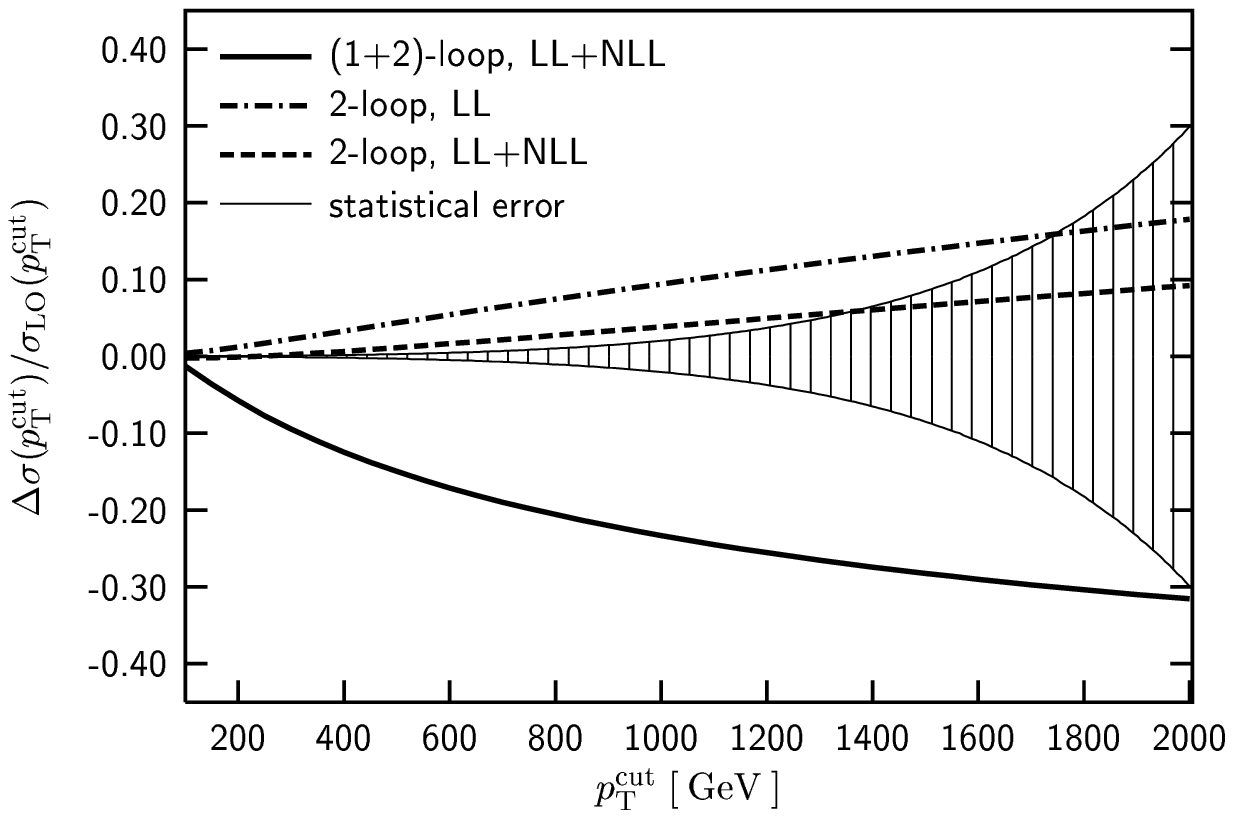, angle=0, width=11.5cm}
\end{center}
\vspace*{-2mm}
\caption{
Relative electroweak correction and statistical error for  
the unpolarised integrated cross section for $ pp \to Z j$ at $\sqrt{s}=14\TeV$
as a function of $\pTcut$:
(1+2)-loop LL+NLL (solid), 2-loop LL
(dash-dotted) and 2-loop LL+NLL (dashed) correction and statistical error
(shaded region) with respect to the lowest order cross section.
}
\label{fig:lhc4}
\end{figure}

The size of the corrections relative to the leading order $\pT$ distribution is shown in Fig.~\ref{fig:lhc1}b. 
These relative corrections increase with $\pT$. 
For high $\pT$ they yield a negative contribution of tens of per cent
to the $\pT$ distribution in the one-loop case
and a positive contribution of several per cent in the two-loop case.
More precisely, for $1\TeV \le \pT \le 2\TeV$, the logarithmic 
one-loop electroweak correction decreases the cross section by 
24\% to 39\%, and the two-loop correction moves the result up by 
3\% to 8\%.
The biggest (positive) contribution to the two-loop result is given by the LL
terms, with NLL terms accounting for about  50\% of the size of the 
LL contribution at high $\pT$ but with an opposite sign. 
This observation demonstrates that the control of NLL effects in the two-loop 
approximation is required to predict the cross section at a level
better than 10\%  for high $\pT$.
We note that for the one-loop result the EM contribution grows with $\pT$ but even at the highest $\pT$ considered it does not exceed
3-4\% of the total electroweak one-loop logarithmic correction
and thus is negligible.

\begin{figure}[h!t]
  \begin{center}
\epsfig{file=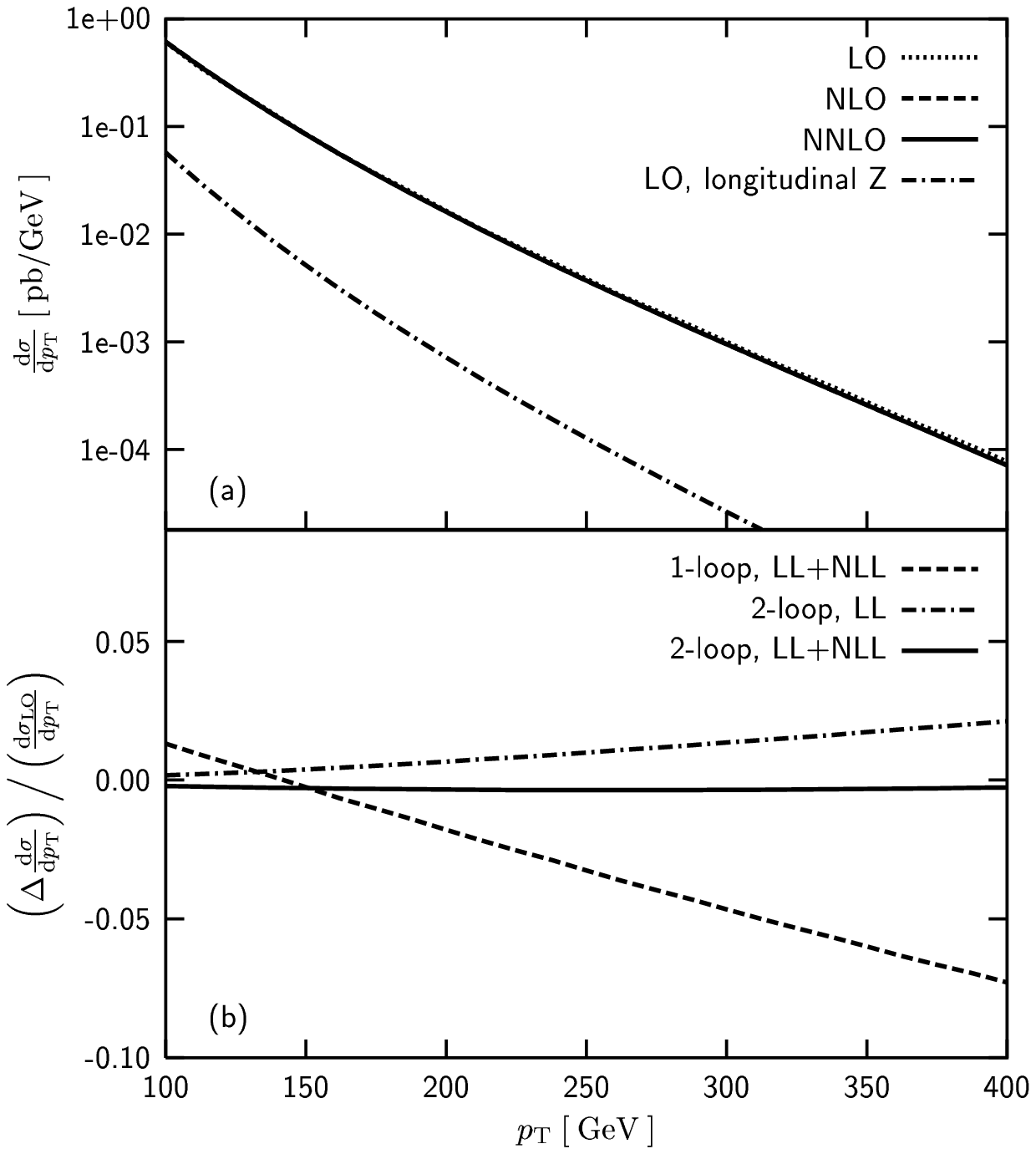, angle=0, width=11.5cm}
\end{center}
\vspace*{-3mm}
\caption{
(a) Transverse momentum distribution for $ p \bar p \to Z j$ at $\sqrt{s}=2\TeV$: 
LO (dotted), NLO (dashed) and NNLO (solid) result for 
unpolarised $Z$ and 
LO contribution from longitudinally polarised $Z$ (dash-dotted).
(b) 
Relative electroweak correction to the unpolarised lowest order $\pT$ distribution for $ p\bar p \to Z j$ at $\sqrt{s}=2\TeV$:
1-loop LLs+NLLs (dashed), 
2-loop LLs (dash-dotted) and 2-loop LLs+NLLs (solid). 
}
\label{fig:tev1}
\end{figure}

The size of the one- and two-loop electroweak logarithmic corrections to the $\pT$ distribution clearly makes them relevant for future measurements 
at the LHC. The significance of the corrections is also illustrated
in Fig.~\ref{fig:lhc4} where we plot the relative one- and two-loop 
corrections to the total cross section obtained by integration over
$\pT$ with $\pT>\pTcut$ as a function of $\pTcut$. 
Additionally, in the same figure we show the statistical
error evaluated assuming  the LHC luminosity and the branching ratio quoted
above. The one-loop correction 
is evidently bigger than the statistical error in the entire
range of $\pT$ considered here.
The two-loop correction is larger than (or comparable to) the anticipated statistical error for values of $\pT$ up to about $1500\GeV$. 
In fact, the correction from the LL term is
significantly larger, being, however, strongly reduced by the
NLL correction. 
This demonstrates that at the two-loop level at least leading- and
next-to-leading logarithms will be relevant for a
reliable prediction.
Of course, a full estimate of the experimental error requires
including systematic effects what leads to an increase in the error value.
Yet, Fig.~\ref{fig:lhc4} indicates relevance of the electroweak
corrections for the precise analysis of experimental data.

We also compared the lowest order $\pT$ distribution and 
one-loop logarithmic correction for the LHC with the results of \citere{Maina:2004rb}.
In their calculations the authors of \citere{Maina:2004rb} introduced rapidity cuts, chose the scale of running $\alpha_\rS$ to be $\mu^2=\shat$,
and while evaluating the one-loop corrections did not consider the 
electromagnetic part but only the weak contributions.
 Taking into account these differences
we find a good agreement for the lowest order $\pT$ distribution and one-loop
corrections, at the level of a per cent for the latter. 
This shows that the main part of the one-loop EW correction is given by
the logarithmic contribution.

\begin{figure}[t]
  \begin{center}
\epsfig{file=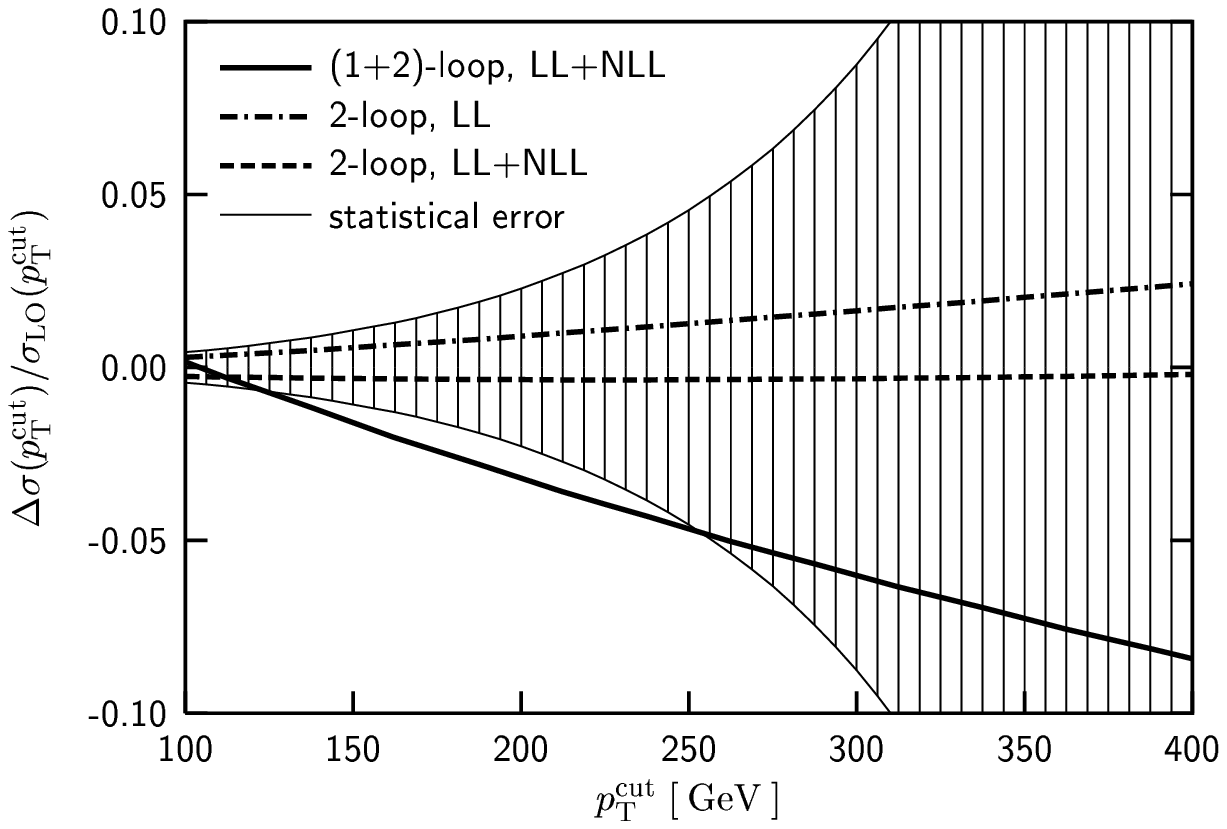, angle=0, width=11.5cm}
\end{center}
\vspace*{-3mm}
\caption{
Relative electroweak correction and statistical error for  
the unpolarised cross section for $ p\bar p \to Z j$ at 
$\sqrt{s}=2\TeV$ as a function of $\pTcut$: 
(1+2)-loop LL+NLL (solid), 2-loop LL
(dash-dotted) and  2-loop LL+NLL (dashed) correction and statistical error
(shaded region) with respect to the lowest order cross section.
}
\label{fig:tev4}
\end{figure}

Results for the $Z+1$ jet production at the Tevatron are shown in
Fig.~\ref{fig:tev1}a,b and in Fig.~\ref{fig:tev4} for the $\pT$
distribution and the integrated cross section, respectively. 
Contrary to 
the LHC case, at the Tevatron the one- and two-loop logarithmic 
electroweak corrections to the cross section do not bear 
much significance for the precise measurement, as illustrated
 in Fig.~\ref{fig:tev4}.
Comparing our results with those of \citere{Maina:2004rb}
we note a disagreement for the lowest order $\pT$ distribution by roughly a
factor of ten at high $\pT$ values. 

\section{Conclusions}

Experiments at the LHC will for the first time explore energies 
in the range well beyond \mbox{1 TeV}. In this region, where $\shat\gg M_{W}^2\sim M_{Z}^2$, the electroweak radiative corrections become important due to large logarithms of $\shat/M_W^2$.
In the present paper we have
studied their impact on hadronic $Z$-boson production at large transverse momentum. 
We have calculated the one- and two-loop 
electroweak corrections in logarithmic approximation,
including leading and next-to-leading logarithmic terms.
These corrections increase rapidly with $\pT$. 
At the LHC, for values of $\pT$ beyond 1 TeV
the one-loop terms amount to several tens of per cent, reaching up to 40
per cent at 2 TeV. These results are in good agreement with those based
on an explicit one-loop calculation \cite{Maina:2004rb}. To fully control the
prediction with a precision that may  be reached in the final stage of
LHC operation, even the dominant two-loop terms must be included.
Significant compensations between leading and next-to-leading two-loop logarithms are observed. The combined two-loop effect of several per cent is comparable to the anticipated statistical error at the LHC. 
Contrary to the LHC, at the Tevatron the one- and two-loop logarithmic electroweak corrections do not have numerical significance.

\section*{Acknowledgements}
This work was supported in part by the Deutsche Forschungsgemeinschaft 
in the Sonderforschungsbereich/Transregio SFB/TR-9 ``Computational
  Particle Physics'' 
and by  BMBF Grant No. 05HT4VKA/3.
A.~K. would also like to acknowledge financial 
support from the Graduiertenkolleg "Hochenergiephysik und Teilchenastrophysik".

\end{document}